\documentclass[trackchanges]{aastex701}

\begin{document}

\title{Fine Structure and Formation Mechanism of a Sunspot Bipolar Light Bridge in NOAA AR 13663}

\author[0009-0003-1469-236X, sname='Qiao']{Fangfang Qiao}
\affiliation{School of Earth and Space Sciences, Peking University, Beijing 100871, People's Republic of China}
\affiliation{State Key Laboratory of Solar Activity and Space Weather, National Space Science Center, Chinese Academy of Sciences, Beijing 100190, People’s Republic of China}
\email{fangfangqiao@stu.pku.edu.cn}

\author[0000-0001-5612-4457, sname='Li']{Hao Li}
\affiliation{State Key Laboratory of Solar Activity and Space Weather, National Space Science Center, Chinese Academy of Sciences, Beijing 100190, People’s Republic of China}
\email[show]{lihao01@nssc.ac.cn}
\correspondingauthor{Hao Li}

\author[0000-0001-5099-8209, sname='Wang']{Jiasheng Wang}
\affiliation{School of Earth and Space Sciences, Peking University, Beijing 100871, People's Republic of China}
\email{jiasheng.wang@pku.edu.cn}

\author[0000-0001-9491-699X]{Yadan Duan}
\affil{Yunnan Observatories, Chinese Academy of Sciences, Kunming, 650216, China}
\affiliation{Yunnan Key Laboratory of Solar Physics and Space Science, Kunming, 650216, People's Republic of China} 
\email{duanyadan@ynao.ac.cn}

\author[0000-0001-5657-7587, sname='Sun']{Zheng Sun}
\affiliation{School of Earth and Space Sciences, Peking University, Beijing 100871, People's Republic of China}
\affiliation{Leibniz Institute for Astrophysics Potsdam, An der Sternwarte 16, 14482 Potsdam, Germany}
\email{2201112089@stu.pku.edu.cn} 

\author[0000-0002-0786-7307,sname='Li']{Leping Li}
\affiliation{State Key Laboratory of Solar Activity and Space Weather, National Astronomical Observatories, Chinese Academy of Sciences, Beijing 100101, People's Republic of China}
\email{lepingli@nao.cas.cn}

\collaboration{all}{}

\begin{abstract}

Bipolar Light Bridges (BLBs) are bright regions located between sunspot umbrae of opposite magnetic polarity. They are typically characterized by strong magnetic fields and intense flows, which are believed to be closely associated with major solar flares. Despite their importance, their fine structure, formation and evolution remain poorly understood. In this work, we analyze the observations of a well-defined BLB obtained by the Goode Solar Telescope at the Big Bear Solar Observatory and the Helioseismic and Magnetic Imager onboard the Solar Dynamics Observatory. The high-resolution GST observations reveal that the BLB is composed of fine, penumbral filament-like structures with widths of approximately 100–150 km. The corresponding Doppler velocity maps present a stable pattern of spatially adjacent red- and blueshifted patches within the BLB throughout the 5.5-hour GST observation. HMI observations show that the BLB arises from the converging and shearing motions of sunspots with opposite polarities. Penumbral regions originating from different polarities gradually evolve and interact, eventually forming the BLB. The observed Doppler velocity pattern, characterized by red- and blueshifted patches, can be interpreted as a projection effect of the Evershed flow within the penumbrae. Therefore, we argue that the BLB is formed through the compression and stretching of penumbral structures from oppositely polarized sunspots.

\end{abstract}


\section{Introduction}       

\label{sect:intro}
Sunspots are among the most prominent features in the solar atmosphere. Various small-scale and complex phenomena have been reported in the regions above sunspots, making them of great interest in the solar community \citep{2007KatsukawaSci...318.1594K,Hui2014sunspotaApJ...790L..29T,Tian2014,Huisunspotb2014ApJ...786..137T,Ruan2015ApJ...812..120R,deng2016multi,Alpert2016ApJ...822...35A,tiwari2016transition,tiwari2018evidence,Srivastava2018NatAs...2..951S,Chen2022,Yuan2023}. Due to their strong magnetic fields, convective energy transport in sunspots is significantly suppressed, giving rise to their low temperature and dark appearance compared to quiet regions \citep{Solanki2003A&ARv..11..153S,Bushby2008MNRAS.387..698B,Rempel2011LRSP....8....3R,Aparna2025ApJ...987...98A}. The darkest region is referred to as umbra, which is usually surrounded by penumbral, formed of filaments and spines \citep{1953sun..book..532C,2013TiwariA&A...557A..25T,Tiwari2015A&A...583A.119T,Tiwari2017arXiv171207174T,Liqiao2018ApJ...857...21L,HinodeReviewTeam2019PASJ...71R...1H}. In penumbral filaments regions, the radial flow, roughly along the nearly horizontal magnetic field lines, was named Evershed flow after its discovery in 1909 with speeds of several km s$^{-1}$ \citep{1909MNRAS..69..454E}. Magnetohydrodynamic (MHD) simulations \citep{2007ApJ...669.1390H,2009ApJ...691..640R, Rempel2009b,2011ApJ...729....5R,2012ApJ...750...62R, Rempel2011} demonstrated that hot plasma rising from beneath the photosphere is guided by inclined magnetic fields, producing radial outflows along the filaments. Observational studies \citep{2015EstebanPozuelo,joshi2011convective,scharmer2012sst,scharmer2013opposite, Tiwari2013ATiwari&A...557A..25T} further revealed that part of this plasma undergoes lateral overturning and sinks back below the surface, supporting a scenario of overturning convection in elongated convective cells: strong outflows occur along the filament axes, while weaker downflows are present at their edges.

Umbra regions may exhibit intricate bright substructures, including light bridges (LBs), where the weaker and more horizontally oriented magnetic fields allow partial convective motions \citep{Leka1997ApJ...484..900L, Toriumi2015, Wang2018a, Zhang2018ApJ, LiFuyu2024}. LBs appear as bright, elongated features crossing the umbra and sometimes connecting to the penumbra \citep{shimizu2011,liu2012,Kamlah2023}. Various dynamic activities have also been observed above light bridges. Chromospheric observations in the $H\alpha$  and \ion{Ca}{2} passbands frequently reveal long-lasting, recurring surge-like or jet-like events, suggesting that LBs play an important role in supplying energy and mass into the higher solar atmosphere \citep{Shimizu2009ApJ...696L..66S,Bharti2015MNRAS.452L..16B,YuanLb2016A&A...594A.101Y,Hou2022ApJ...929...12H,HuiLB2018ApJ...854...92T}. 
Morphologically, LBs are generally categorized into granular and filamentary LBs \citep{Rimme2008ApJ...672..684R,Kleint2013ApJ...770...74K,Lagg2014A&A...568A..60L,Guglielmino2017ApJ...846L..16G,wang2018ApJ...852L..18W}. \citet{Li2021RAA....21..144L} classified LBs according to their formation processes: penumbral intrusion (type A), sunspot merging (type B), and umbral-dot emergence (type C), and conducted the first statistical study of LB formation.

A particularly interesting type of LBs are bipolar light bridges (BLBs), which occur between two umbral cores of opposite magnetic polarity \citep{Castellanos2025}. BLBs are exclusively found within $\delta$-type sunspot groups, where umbrae of opposite polarity share a common penumbra, and their long axis is typically aligned with the polarity inversion line (PIL). They often contain “superstrong” magnetic fields that exceed typical umbral strengths \citep{Zirin1993,Okamoto2018ApJ...852L..16O,Wang2018,Toriumi2019ApJ...886L..21T,Castellanos2020ApJ...895..129C,Jiayi2023ApJ...955...40L}. Recently, a statistical study by \citet{Castellanos2025} identified 98 BLBs across 51 sunspot groups. By applying the SPINOR code \citep{Frutiger2000A&A}, they revealed a lower limit of 4.5 kG for the magnetic field strength of BLBs at optical depth unity. The origin of such strong horizontal magnetic fields had been investigated through radiative MHD simulations by \citet{Hotta2020MN}, who successfully reproduced a $\delta$-type sunspot structure, including a BLB with magnetic field strengths exceeding 6 kG. Their results suggest that shear motions, driven by the rotational movement of two sunspots originating from the deep convection zone, amplify the magnetic fields within BLBs to superequipartition levels\citep{Hotta2020MN}.

Another interesting phenomenon is the vigorous photospheric plasma motion associated with BLBs (or orphan penumbrae lacking surrounding umbrae), manifested as spatially adjacent red- and blueshifted Doppler patches \citep{Zuccarello2014,Cristaldi2014,Jiayi2023ApJ...955...40L}. The longitudinal Doppler velocities ranging from a few km s$^{-1}$ 
\citep{Takizawa2012,Shimizu2014P,Jaeggli2016} to supersonic speeds (e.g., 14 km s$^{-1}$
; \citeauthor{Martinez1994}, \citeyear{Martinez1994}) are obtained from Spectropolarimetric measurements. These adjacent red and blueshifted patches Doppler signatures may arise from rapid magnetic flux emergence or cancellation, or they can be interpreted as projections of field-aligned flows, such as the Evershed flow \citep{Lites2002,Jaeggli2016,Cristaldi2014}.

The interplay between strong magnetic fields and high-velocity plasma flows can inject substantial magnetic energy and helicity into the corona, thereby facilitating the formation of magnetic flux ropes or filaments and enhancing the eruptive potential of active regions \citep{Welsch2009,sun2024A&A...686A.148S,2025SunarXiv250711790S,Zheng2025SCPMA..6879611Z}. For instance, NOAA AR 12673 hosted a BLB that exhibited adjacent red and blueshifted patches Doppler signatures \citep{Jiayi2023ApJ...955...40L}, and subsequently produced several major flares, including the strongest X9.3 event, which suggests a possible link between the presence of BLBs and large-scale energy release processes \citep{Yang2017,Yan2018ApJ...856...79Y,Getling2019,Moraitis2019,Liu2019ApJ...884...45L}. 

Despite these findings, the fine structure, the formation, and evolution of BLBs are still poorly understood. Some evidence suggests that BLBs may result from the coalescence of opposite-polarity sunspots \citep{Zirin1993}. However, detailed studies on the exact formation process and the mutual interactions remain scarce. In this work, we report a BLB observed by the high-resolution Goode Solar Telescope (GST) at the Big Bear Solar Observatory (BBSO). Its host active region was continuously monitored by the Helioseismic and Magnetic Imager, allowing us to investigate its formation and evolution. This dataset offers a unique opportunity to investigate the BLB’s fine structure, formation process, and dynamic behavior in detail. The outline of this paper is as follows.
Section \ref{sect:Obs} describes the observational data. The results are presented in Section \ref{sect:Res}. Section \ref{sect:discussion} discusses caveats and possible flow-driving mechanisms, and conclusions are given in Section \ref{sect:conclusion}

\section{Observations}
\label{sect:Obs}
The NOAA Active Region (AR) 13663, hosting the target BLB, first appeared on the solar disk on April 30, 2024, at heliographic coordinates N25E34. It remained visible until May 11 2024, when it rotated off the western limb near N27W$0^*$. During its disk passage, NOAA AR 13663 produced 39 C-class, 35 M-class, and 5 X-class flares. The most energetic event was an X4.52 flare, which peaked at 06:35 UT on May 6 2024.

Imaging observations of this active region in the TiO and $H\alpha$ bands were obtained with the GST/BFI and VIS, respectively. The TiO band filter is centered at 705.7 nm with an approximate bandwidth of 1 nm, and an exposure time of 0.8 ms. The pixel scale after speckle reconstruction is about 0.030$^{\prime\prime}$, enabling characterizing the fine structure of the BLB. Despite intermittent data gaps, the observations of this active region in TiO band continuously covered around 5.5 hours starting at 16:38 UT on May 3, 2024. During the 5.5-hour observation period, the BLB structure remained remarkably stable. Therefore, a representative time frame was selected for detailed analysis in this study. Narrowband $H\alpha$  spectral scans were recorded by the GST/VIS instruments using a single Fabry–P{\'e}rot etalon, with a 0.07 ~\AA\ bandpass ranging from 550 to 700 nm and a pixel scale of 0.029$^{\prime\prime}$ per pixel.

Spectropolarimetric observations of the BLB region were performed with GST/NIRIS \citep{NIRIS2012}. This Fabry–P{\'e}rot-based imaging system recorded full Stokes parameters covering a wavelength of 3.16  ~\AA\ across the \ion{Fe}{1} spectral line at 1560 nm with a spectral sampling of 0.158~{\AA}. The field of view of the instrument is 85$^{\prime\prime}$ with a spatial resolution of 0.083$^{\prime\prime}$ per pixel. Milne–Eddington inversion was applied to this data set to retrieve the magnetic field vector and the line-of-sight (LOS) velocity (see Appendix~\ref{ME} for the details).

The continuous observations of this active region was obtained by the Helioseismic and Magnetic Imager \citep[HMI][]{HMIScherrer2012} onboard the Solar Dynamics Observatory \citep[SDO][]{Pesnell2012SDO}, covering its evolution throughout its disk passage. 

HMI scans the Stokes parameters of the \ion{Fe}{1} 6173~\AA\ line with a 75 m\AA\ narrow filter at six wavelength positions. The spatial resolution of the observation is approximately 0.504$^{\prime\prime}$ per pixel. The magnetic field vector and LOS velocity are produced with the VFISV code \citep{Borrero2011SoPh}. In this work, we primarily used the continuum intensity images, Dopplergrams, and line-of-sight magnetograms. The continuum intensity images are normalized to the quiet Sun continuum intensity ($I_{qs}$), defined as the mean intensity within a quiet Sun region. The umbra is identified as the area where the continuum intensity falls below 0.55\,$I_{c}$. Doppler velocities were calibrated by assuming that the plasma within the regions with intensity below 0.5\,$I_{c}$ is at rest. This reference minimizes the impact of convective blueshifts in the quiet Sun by treating the umbral region as stationary \citep{Dravins1981A&A....96..345D}. The PIL shown in Figures~\ref{figure:fig2} is determined from the longitudinal magnetic fields. Specifically, if a $3\times3$ pixel region contains both positive ($> 15~\mathrm{G}$) and negative ($< -15~\mathrm{G}$) longitudinal magnetic fields, the center pixel is considered as part of the PIL. Since this active region was located as a heliocentric angle of 30 degrees, the determination of the PIL suffers from projection effects.

To align data from different channels accurately, we used the SIFT algorithm implemented in Python’s OpenCV library. SIFT detects and matches keypoints robustly across scale and rotation differences, enabling precise multi-channel registration \citep{lowe2004distinctive}.

\section{Results}
\label{sect:Res} 

\begin{figure*}[ht!]
\centering
\begin{interactive}{animation}{animation1.mp4}
\includegraphics[width=1\textwidth]{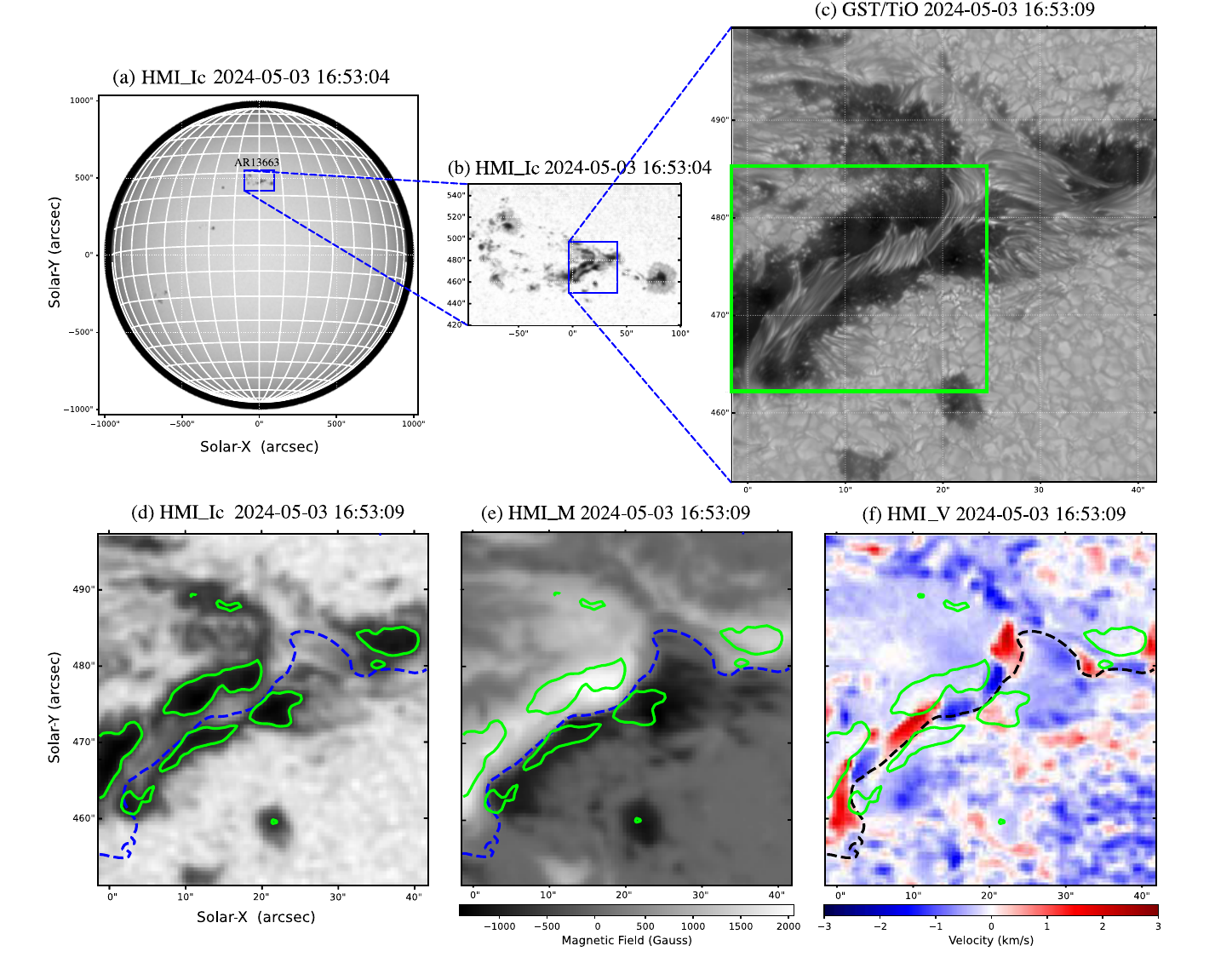}
\end{interactive}
\caption{Overview of the BLB at 16:53 UT on 2024 May 3.
(a) Full-disk solar image with NOAA AR 13663 marked by a blue rectangle, corresponding to the FOV shown in panel (b).
(b) Zoomed-in view of NOAA AR 13663; the blue rectangle indicates the FOV shown in panels (c)–(f).
(c) High-resolution GST/TiO image of the light bridge. The lime rectangle shows the FOV of Figure~\ref{figure:fig2}. Panel (c) is available as an animation showing the BLB over the entire 5.5-hour interval starting at 16:38 UT on 3 May 2024. The real-time duration is 16 seconds.
(d)–(f) SDO/HMI observations of the same region, corresponding to the continuum intensity, LOS magnetogram, and LOS Doppler velocity, respectively. Lime contours denote the regions with continuum intensity $< 0.35$ $I_\mathrm{c}$. Blue and black dashed lines indicate locations where the line-of-sight magnetic field is zero.
\label{figure:fig1}}
\end{figure*}

Figure~\ref{figure:fig1} provides an overview of the BLB region at 16:53 UT on 2024 May 3. By this time, NOAA AR 13663 had developed into a mature sunspot group, spanning from 420$^{\prime\prime}$ to 550$^{\prime\prime}$ in solar latitude and positioned near the central meridian in longitude, as shown in Figure~\ref{figure:fig1}(a). An enlarged view of NOAA AR 13663 is presented in Figure~\ref{figure:fig1}(b), where the blue rectangle indicates the field of view (FOV) corresponding to Figures~\ref{figure:fig1}(c)–(f). The BLB is located at the center of the active region, while a relatively simple sunspot with well-developed peripheral penumbrae lies to its west. 

Figures~\ref{figure:fig1}(d) and (e) show the HMI continuum intensity and LOS magnetogram, respectively, indicating that the BLB is located above the zero-crossing of the LOS magnetic field, as outlined by the blue and black dashed lines, with most of its structure situated in the region of positive polarity.
This apparent location may be influenced by projection effects, leading to a systematic shift in the zero-crossing.

Due to the low spatial resolution of SDO/HMI, the BLB appears as a narrow bright structure separating the two umbral cores of the $\delta$-configuration sunspot, as indicated by the combined analysis of Figures~\ref{figure:fig1}(d) and (e). The limited spatial resolution makes it difficult to resolve its fine structure using HMI data alone.

High-resolution TiO images from GST (Figure~\ref{figure:fig1}(c)) provide a more detailed view of the BLB, which consists of filamentary structures connecting the positive and negative umbral cores.

Segments of filamentary structures within the BLB that are aligned more parallel to the umbral boundaries tend to be darker in TiO images, whereas those oriented more tangentially appear brighter, particularly in the central region of the BLB. The LOS Doppler velocity map in Figure~\ref{figure:fig1}(f) shows both upflows (blueshift) and downflows (redshift) in the BLB region, primarily distributed along the northern segment of the PIL.  
 
Throughout the GST observing period from 16:38 UT to 22:05 UT on 2024 May 3, the filamentary structures shown in the animation of Figure~\ref{figure:fig1}(c) maintained a stable morphology.

\subsection{Fine Structure}
\label{sub:fine structure}
\begin{figure*}[ht!]
\centering
\begin{interactive}{animation}{animation2.mp4}
\includegraphics[width=1\textwidth]{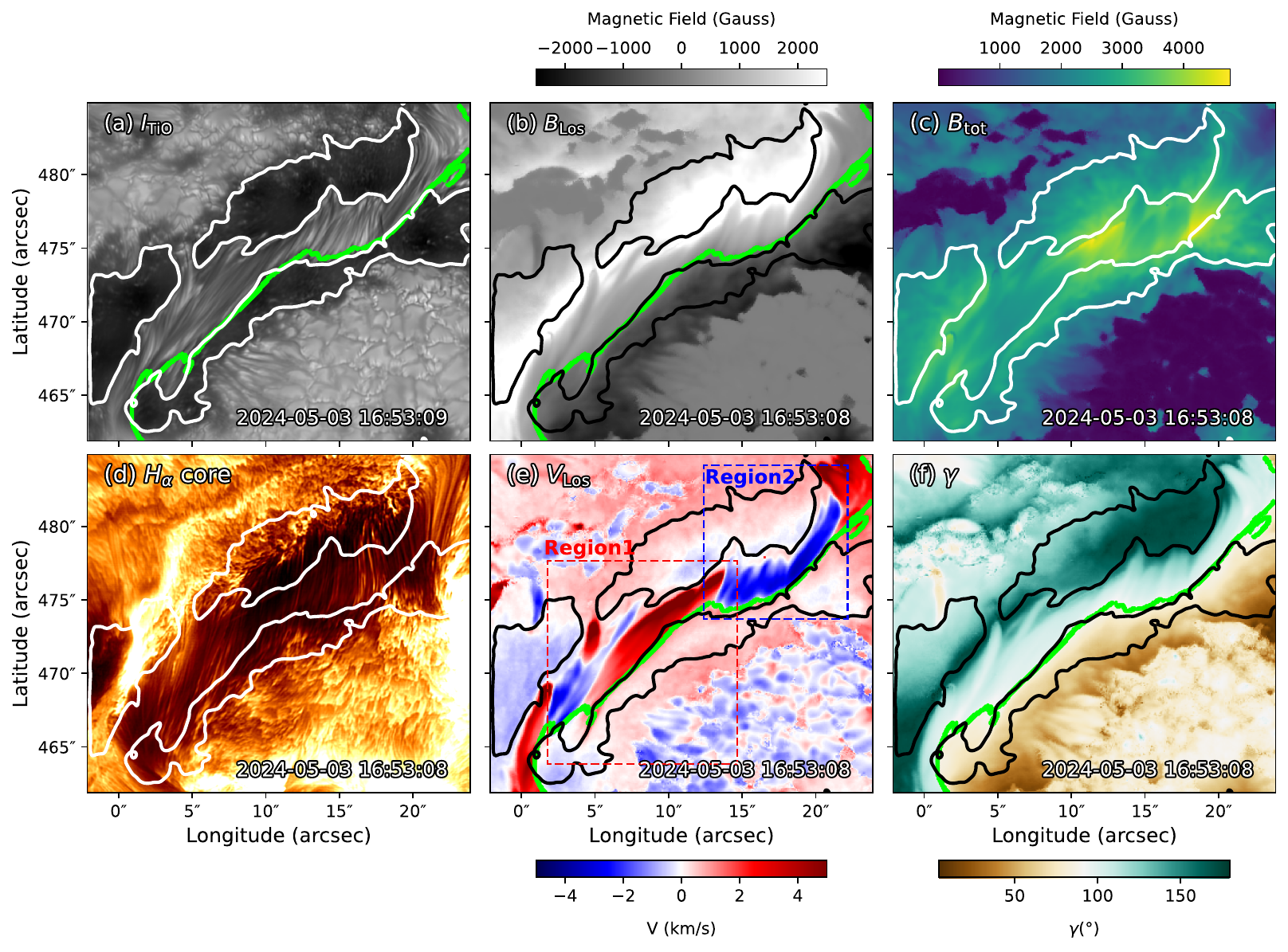}
\end{interactive}
\caption{High-resolution observations of the BLB using BBSO/GST, focusing on the region of interest (ROI) shown in Figure~\ref{figure:fig1}(c).
(a) GST/BFI TiO image.
(b)-(c): Line-of-sight (LOS) magnetic field and total 
magnetic field derived from GST/NIRIS observations. 
(d) GST/VIS H$\alpha$ line center. 
(e)-(f): Doppler velocity, and horizontal magnetic field 
derived from GST/NIRIS observations. 
Red dashed box Region1 and blue dashed box Region2 in (e) mark the FoV in Figure~\ref{figure:fig3} and Figure~\ref{figure:fig4}, respectively.
White contours and black contours in (a)-(f) denote regions where the continuum intensity is below $0.2\space I_\mathrm{c}$ of intensity of TiO. The green curves in (a), (b), (e), and (f) mark the PIL.
See Section~\ref{sub:fine structure} for details. An animation showing the evolution of BLB based on NIRIS observations is available. The animation shows the sequence from May 3rd, 2024 from 16:40 to 21:40. The real-time duration is 1.33 seconds.
\label{figure:fig2}}
\end{figure*}

\begin{figure*}[ht!]
\centering
\begin{interactive}{animation}{animation3.mp4}
\includegraphics[width=1\textwidth]{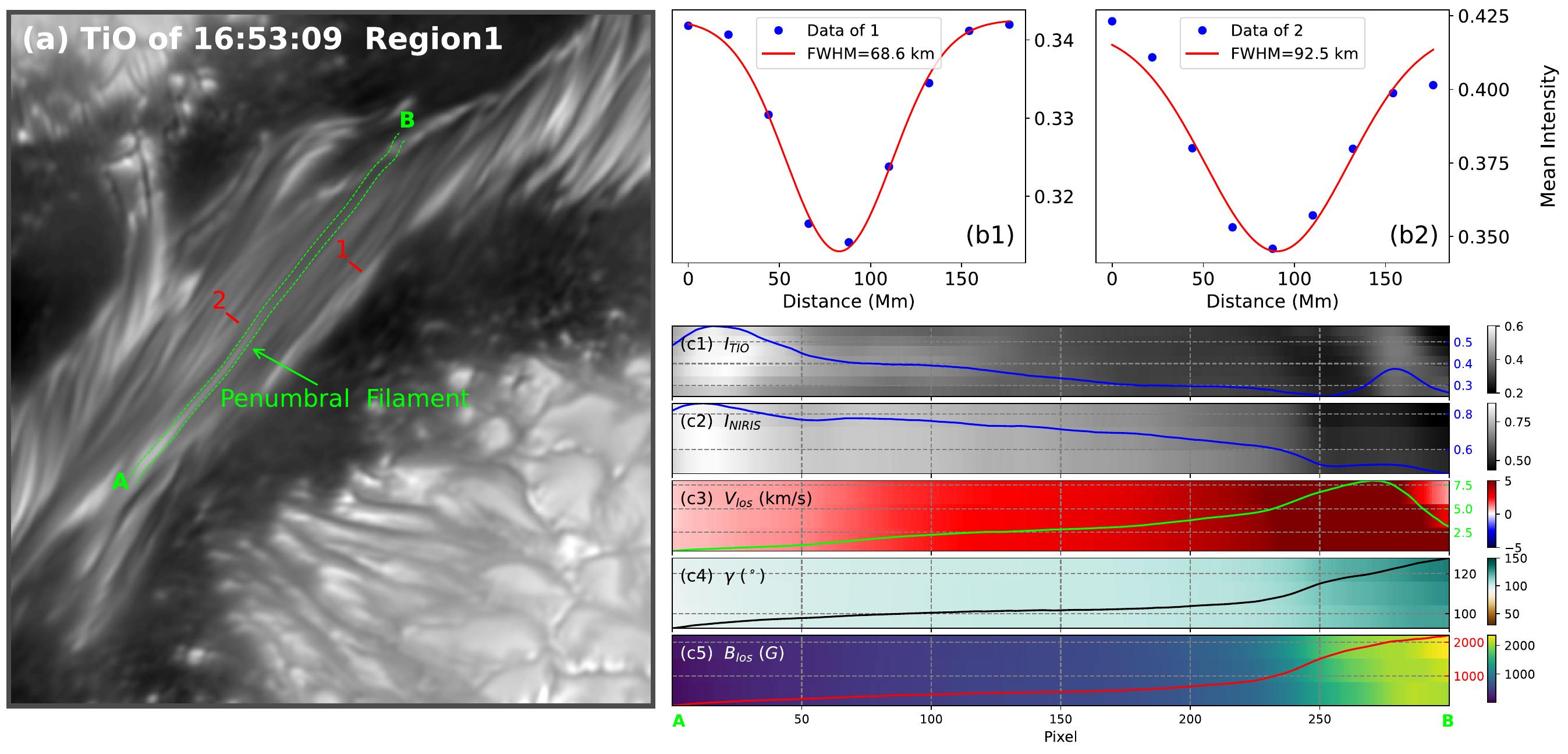}
\end{interactive}
\caption{Properties of Region 1(redshifted Doppler region) shown in Figure~\ref{figure:fig2}(e).
(a) Image in the TiO band. The short red lines labeled 1 and 2 indicate the positions where the filament widths are measured in panels (b1) and (b2), respectively. Panel (a) is available as an animation showing the evolution of the redshifted region of the BLB. The animation runs from 16:37:09 to 17:39:09 with a real-time duration of 8 seconds. The lime dashed curved lines mark a relatively representative penumbral filament corresponding to the NIRIS data shown in panels (c1)–(c5); the label A–B indicates the orientation.
(b1)–(b2) Gaussian fitting of the intensity variation at locations 1 and 2 in panel (a), respectively.
(c1)–(c5) Maps within the lime dashed curved lines in panel (a), including the TiO intensity, NIRIS intensity, inferred LOS velocity, magnetic inclination, and LOS magnetic field strength. The spatially averaged value is overlaid on each map as a solid curve.}
\label{figure:fig3}
\end{figure*}

\begin{figure*}[ht!]
\centering
\begin{interactive}{animation}{animation4.mp4}
\includegraphics[width=1\textwidth]{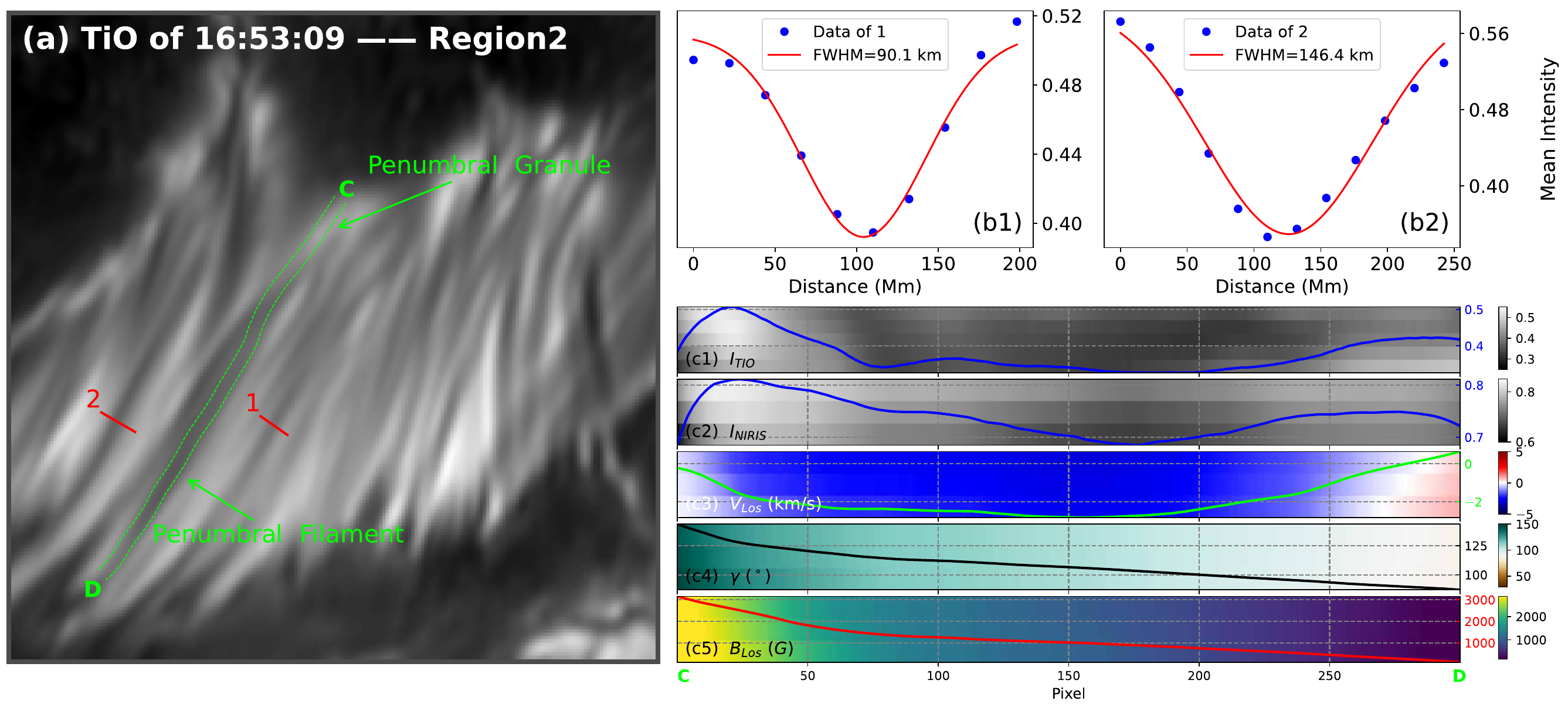}
\end{interactive}
\caption{
Same as Figure~\ref{figure:fig3}, but for Region~2 (the blueshifted Doppler region). Panel (a) is available as an animation showing the evolution of the blueshifted region of the BLB. The animation runs from 16:37:09 to 17:39:09 with a real-time duration of 8 seconds. 
Label C--D in panel (a) indicates the orientation of a representative penumbral filament, and point C marks the location of the penumbral grain.
\label{figure:fig4}}
\end{figure*}

The high-resolution GST observations enable a detailed study of the fine-scale structure of the BLB. Figure~\ref{figure:fig2} presents a zoomed-in view of the BLB’s region, corresponding to the lime rectangle in Figure~\ref{figure:fig1}(c). As shown in Figure~\ref{figure:fig2}(a), the GST/TiO image reveals that the BLB consists of multiple filamentary segments, each exhibiting a distinct orientation, which is closely similar to the structures seen in the MHD simulation by \citet{Hotta2020MN}. The H$\alpha$ line-center image (Figure~\ref{figure:fig2}(d)) shows that the overlying chromospheric layer is dominated by arching loops, which almost entirely obscure the BLB. The total magnetic field map derived from GST/NIRIS inversions (Figure~\ref{figure:fig2}(b)) confirms the findings of \citet{Castellanos2025}, showing that the strongest magnetic fields are concentrated along the BLB’s edge rather than within the umbra. Consistent with SDO/HMI observations, the GST/NIRIS inversion results (Figure~\ref{figure:fig1}(d) and (e)) show that the BLB is predominantly located within the positive polarity region and exhibits strong, large-scale adjacent red- and blueshifted Doppler patches.
Compared to the low-resolution HMI observation, these measurements reveal intricate structures in both the LOS magnetic field and Doppler velocity, exhibiting well-defined filamentary features similar to the radiative structures of the BLB shown in Figure~\ref{figure:fig2}(a). 
Moreover, as shown in Figure~\ref{figure:fig2}(f), the magnetic inclination within the BLB is approximately 90$^\circ$, indicating that the magnetic field is oriented nearly parallel to the solar surface. 
We also examine the temporal evolution of the BLB. Throughout the GST observing period, the large-scale flow pattern in the BLB region remains stable, as shown in the accompanying animation. Although the animation displays an apparent signal near the edges of the field of view, this effect is caused by the image rotation correction and does not affect the central structures that are the focus of our analysis.

A key characteristic of the BLB is the persistent spatially adjacent red- and blueshifted Doppler patches. For detailed analysis, we focus on two regions marked by red and blue dashed rectangles in Figure~\ref{figure:fig1}(e), with Region 1 corresponding to redshifted areas and Region 2 to blueshifted areas. 
As shown in Figures~\ref{figure:fig3} and \ref{figure:fig4}, our analysis focuses on two aspects: (1) measuring the widths of dark filaments, and (2) selecting a representative filament with a morphology similar to penumbral filaments to investigate spatial correlations among physical parameters inferred from NIRIS. For better comparison, the spatially averaged value is overlaid on each map as a solid curve in Figures~\ref{figure:fig3}(c1)-(c5) and \ref{figure:fig4}(c1)-(c5).

Figure~\ref{figure:fig3} presents the redshifted region 1. The widths of Filaments were measured using a Gaussian fit to the intensity variation along the red line shown in Figure~\ref{figure:fig3}(a). As shown in Figure~\ref{figure:fig3}(b), the full width at half max (FWHM) for the two selected regions are 68.6 km (location 1) and 92.5 km (location 2), respectively. 
Within the region marked by the lime dashed curved lines, the intensity from TiO and NIRIS, the LOS velocity and the magnetic field parameters are rotated and displayed horizontally in Figure~\ref{figure:fig3}(c1)-(c5). 
Along the A–B direction, a representative filament is selected that closely resembles typical penumbral filaments as depicted by \cite{Tiwari2013ATiwari&A...557A..25T}, exhibiting a bright head at point A, a darker main body, and a relatively bright tail at point B in the TiO intensity.
The spatial resolution of NIRIS is lower than that of TiO. As a result, a very precise correspondence cannot be clearly identified in the intensity maps in Figure~\ref{figure:fig3}(c1) and (c2). Nevertheless, a brighter head and a darker main body can still be distinguished, and the tail also shows a slight intensity enhancement. As shown in Figure~\ref{figure:fig3}(c3), the main body of the filament exhibits redshifted Doppler velocities along the A–B direction. The redshift is not spatially uniform along the filament: it is close to zero near the filament head, while it increases significantly toward the tail. In addition, the magnetic field in the main body of the filament is relatively horizontal and weak which are shown in Figure~\ref{figure:fig3}(c4) and (c5). As shown in the animation, the filament displays a clear material flow directed from the A end to the B end. 

The blueshifted Region 2 is shown in Figure~\ref{figure:fig4}.
As illustrated in Figure~\ref{figure:fig4}(b), we select two locations to measure the filament widths, which are 90.1~km and 146.4~km, respectively. We choose the filament C–D as a representative example for detailed analysis. This filament also exhibits a bright head, a relatively faint main body, and a comparatively bright tail. Along the C–D direction, the main body of the filament is characterized by blueshifted Doppler velocities, which are more uniform along the filament compared to those in Region~1. The magnetic field in the filament main body is relatively horizontal and weak, as shown in Figures~\ref{figure:fig4}(c4) and (c5). The filamentary structures in this region can be broadly classified into two types. The left part, where the representative filament was selected and measured, appears brighter, whereas the right part exhibits filamentary structures more similar to those in Region~1. In the right part, the animation reveals a clear material flow directed from the southern toward the northern portion of the region. In contrast, in the left part, near the filament head (the C end), the bright plasma is observed to move toward the northern umbra.

The close connections between the structural morphology, flow dynamics, and Doppler velocities observed in these BLBs and those in typical sunspot penumbrae will be further explored in the Section~\ref{sect:discussion}.

\subsection{Formation of the BLB}
\label{sub:Origin}

\begin{figure*}[ht!]
\centering
\begin{interactive}{animation}{animation5.mp4}
\includegraphics[width=1\textwidth]{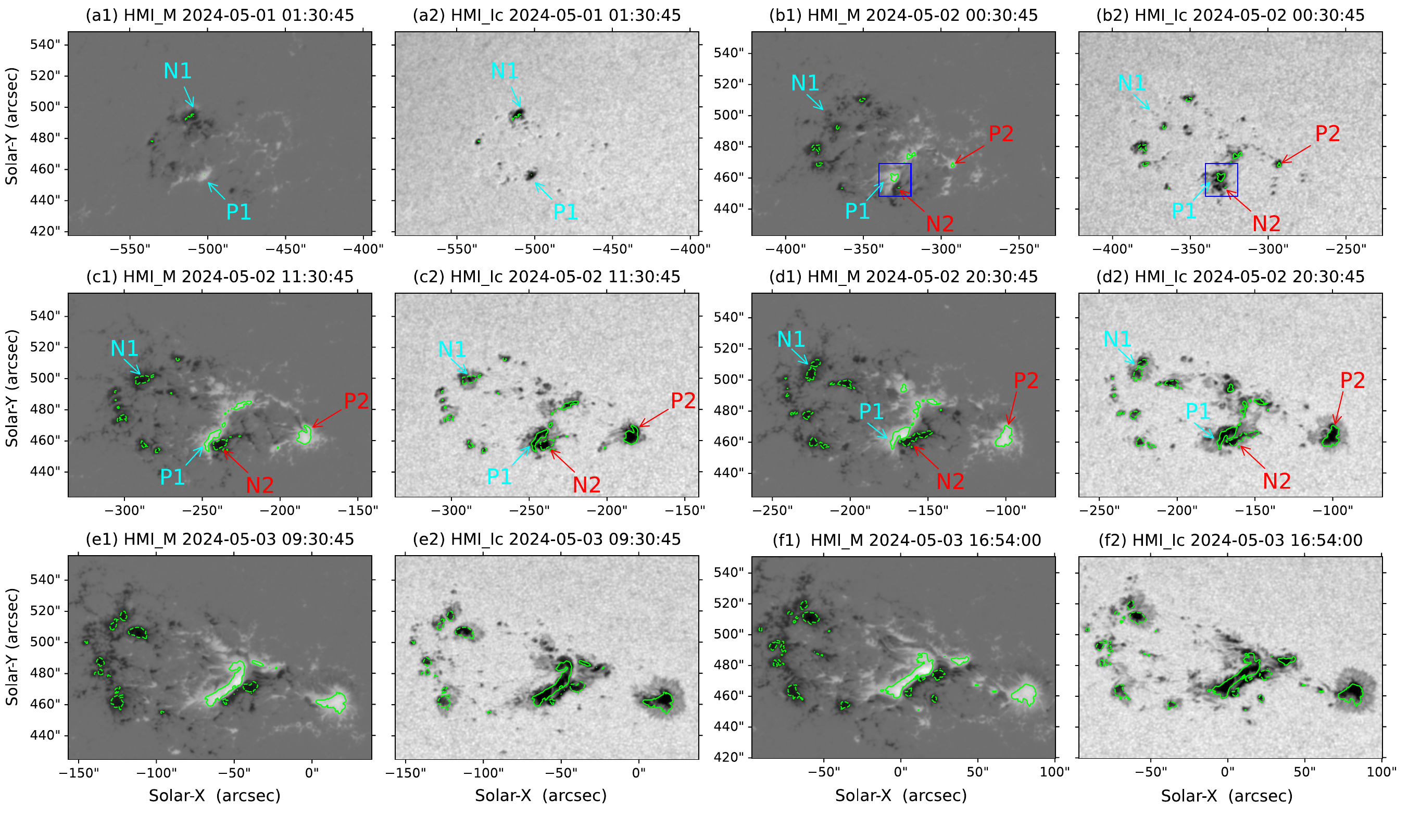}
\end{interactive}

\caption{Evolution of NOAA AR 13663.
(a1)–(f1) SDO/HMI LOS magnetograms from 01:30 UT on 1 May to 16:54 UT on 3 May, 2024.
(a2)–(f2) SDO/HMI continuum intensity images for the times as same as panels (a1)–(f1). 
Solid and dashed lime contours denote the LOS magnetic field strengths of $\pm$ 1000 G, respectively.
Blue rectangles in panel (b1) and (b2) mark the ROI shown in Figure~\ref{figure:fig6}.
Cyan arrows labeled N1 (negative polarity) and P1 (positive polarity) mark the pre-exsiting sunspot pair, while red arrows labeled N2 (negative polarity) and P2 (positive polarity) indicate the emerging sunspot pair. 
An animation showing the evolution of NOAA AR 13663 is available. The animation runs from April 29th, 2024 20:29:49 to May 5th, 2024 23:29:05. The animation's real-time duration of 9 seconds.
\label{figure:fig5}}
\end{figure*}

\begin{figure*}[ht!]
\centering
\begin{interactive}{animation}{animation6.mp4}
\includegraphics[width=1\textwidth]{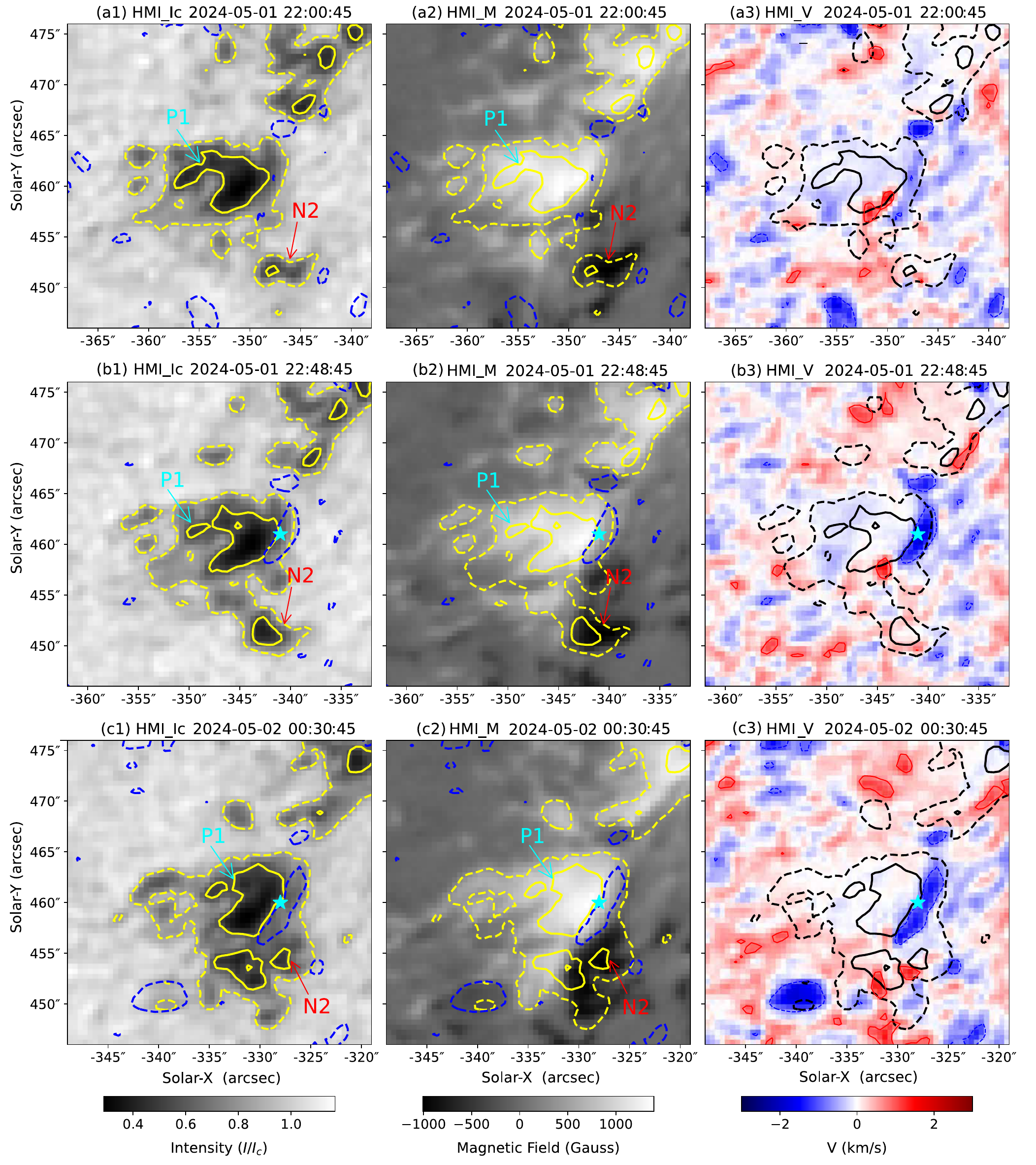}
\end{interactive}
\caption{Evolution of the ROI marked in Figure~\ref{figure:fig5}(b1) and (b2).
(a1)–(c1): SDO/HMI continuum intensity. 
(a2)–(c2): SDO/HMI LOS magnetograms. 
(a3)–(c3): LOS Doppler velocity.
Solid and dotted yellow (black) contours correspond to intensities of 0.55 and 0.85 $I_{c}$, respectively.  
Blue dashed and red solid contours correspond LOS Doppler velocities of $\mp 0.8$ km\,s$^{-1}$, respectively. 
The labels P1 and N2 are same as those in Figure~\ref{figure:fig5}(b1) and (b2).
Cyan stars highlight the penumbral structures of interest.
An animation showing the evolution preceding the close approach of N1 and P2 is available. The animation runs from May 1st, 2024 21:44:49 to May 2nd, 2024 01:29:04. The animation's real-time duration of 6 seconds. 
\label{figure:fig6}}
\end{figure*}

In order to explore the formation of the BLB and the physical processes driving the large-scale flow, we analyzed long-duration time series observations from SDO/HMI. 
Figure~\ref{figure:fig5} illustrates the evolution of NOAA AR 13663 prior to the stable phase of the BLB. 
The sequence illustrates the development of a quadrupolar sunspot group, arising from the emergence of a new sunspot pair (N2–P2) near a pre-existing pair (N1–P1).
As shown in Figure~\ref{figure:fig5}(a1) and (a2), the sunspot pair N1–P1 exists at the early stage, with the P1 migrating westward and N1 migrating eastward.  
The pair N2–P2 subsequently emerges to the west of N1–P1, as shown in Figure~\ref{figure:fig5}(b1) and (b2).
This newly emerged pair exhibits a similar migration pattern, with P2 moving westward and N2 moving eastward.
Consequently, the converging motions of N2 and P1 bring them into interaction.

As N2 and P1 move closer and grow larger, they simultaneously develop penumbrae and gradually evolve into mature sunspots, as shown in Figure~\ref{figure:fig5}(c1) and (c2). 
The region between them exhibits intermediate intensity relative to the quiet Sun and the dark umbrae, while the LOS magnetogram shows relatively weak magnetic fields—both characteristic of a penumbral structure. This area evolves more rapidly than N2 itself, extending northwestward, as clearly illustrated in Figure~\ref{figure:fig5}(d1) and (d2). Figure~\ref{figure:fig5}(e1) and (e2) show that the penumbral area between N2 and P1 is compressed as a result of their mutual motion and shear, consequently forming the BLB.
The BLB subsequently becomes increasingly elongated and west–east aligned, as seen in Figure~\ref{figure:fig5}(f1) and (f2), possibly due to the sunspot rotation. 
These sequent observations indicate that the BLB is formed by the interleaving of penumbral regions originating separately from N2 and P1 as they approached each other.

To further support this conclusion, we examined the evolution preceding the close approach of N1 and P2, focusing on the region outlined by the blue rectangle in Figures ~\ref{figure:fig5}(b1) and (b2).
As shown in Figure~\ref{figure:fig6}(a1)–(a3), before the close interaction of P1 and N2, penumbral structures outlined by dashed yellow and black contours corresponding to an intensity of 0.85 $I_{c}$ already existed around P1. 
Since no clear Evershed flow is detected in the LOS Doppler velocity at this stage, suggesting that the penumbra remains in an early or developing phase. 
Approximately two hours later (Figure~\ref{figure:fig6}(b1)–(b3)), P1 and N2 had moved closer. 
A distinct upflow region, marked by cyan stars, emerges with a LOS velocity exceeding 0.8 km $s^{-1}$. 
At this time, the penumbral segment originating from P1 lies in the northeastern part of the solar disk. 
The observed blueshift is therefore consistent with the radially outward Evershed flow projected along the line of sight, providing further evidence that this structure is part of the penumbra. 
This penumbral structure persists and gradually elongates due to the approaching and shear motions of P1 and N2, as shown in Figure~\ref{figure:fig6}(c1)–(d1). 
Despite the close compression of portions of P1 and N2, the penumbral structure remains stably situated between them, ultimately forming the initial segment of the subsequently observed BLB.

\section{Discussion}
\label{sect:discussion}

\begin{figure*}[ht!]
\centering
\includegraphics[width=1\textwidth]{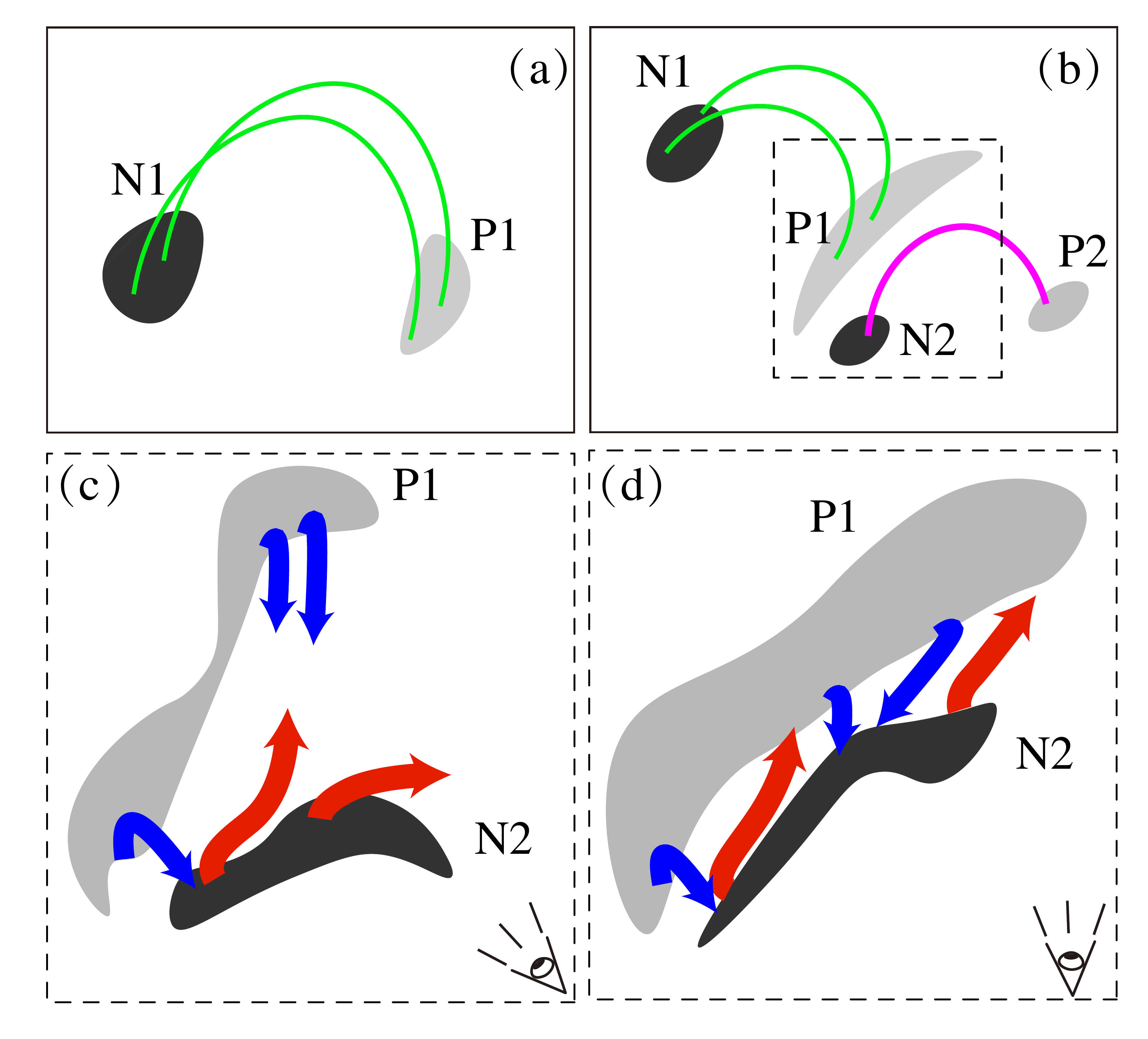}
\caption{Schematic diagrams illustrating the formation of the BLB. Black (N1/N2) and gray (P1/P2) ellipses represent negative and positive magnetic polarities, respectively. In (a) and (b), the green and pink curves illustrate magnetic field lines, connecting the negative and positive magnetic polarities of the two sunspot pairs, respectively. 
Black dashed box in panel (b) marks the ROI, where the BLB forms. 
Broad curved blue and red arrows in panels (c) and (d) denote penumbral flows of P1 and N2, exhibiting Doppler blueshifts and redshifts, respectively. 
Eye symbols are used to denote the observers.
   \label{figure:fig7}}
\end{figure*}

By combining high-resolution GST observations with long-duration SDO/HMI data, we investigate the physical properties of a BLB and propose a possible formation mechanism.

The high-resolution images reveal several defining features of the BLB, as shown in Figure~\ref{figure:fig2}, including including fine-scale filamentary structures, intermediate intensity between the umbra and the quiet Sun, and predominantly horizontal magnetic fields. We also measured the widths of the filaments, which range from 50 to 120~km. One proposed explanation for the origin of these structures involves the flute (or interchange) instability, which can lead to the corrugation of magnetic flux tube boundaries \citep{Spruit2010,Schlichenmaier2016}. This instability arises from perturbations at the flux tube surface and can significantly deform its geometry. The magnetic flux tubes in the photosphere, particularly near the optical depth $\tau_{500} \approx 1$, are especially susceptible to this mechanism, providing a plausible explanation for the observed fine-scale filamentation. 

Notably, the BLB exhibits long-lived and spatially coherent LOS Doppler velocity patterns, adjacent red and blueshifted patches (Figure~\ref{figure:fig1}(f) and Figure~\ref{figure:fig2}(e)). These velocity signatures remain stable throughout the entire duration of the GST observations.

Since the Evershed flow is known to be a possible origin of such flows\citep{Zuccarello2014,Liu2019ApJ...884...45L}, and clear mass motion along the filament is visible in the accompanying movie, the observed Doppler patterns may be interpreted as projections of oppositely directed Evershed flows originating from different umbral cores. In other words, the BLB may consist of different segments of penumbral filaments connected to distinct umbral cores.
At 16:53 UT on 2023 May 3 (the time corresponding to Figure~\ref{figure:fig2}), NOAA AR 13663 was located at a heliocentric angle of approximately $30^\circ$ and about $455^{\prime\prime}$ north of the solar disk center, as shown in Figure~\ref{figure:fig1}(a). In this viewing geometry,  the upflows (blueshift) are associated with the plasma motion toward the solar disk center along penumbral filaments extending from the northern umbra, while the downflows (redshift) are associated with the plasma motion away from the solar disk center along penumbral filaments of the southern umbra \citep{Borrero2005A&A...436..333B}. 

As shown in Figure~\ref{figure:fig3}(a), in redshifted region 1, we select a representative filament segment A–B \citep{Tiwari2013ATiwari&A...557A..25T}. This filament is characterized by a relatively bright head, a weaker main body, and a brighter bump at the tail. According to the magnetoconvective model of the penumbra, hot plasma initially rises near point A and is subsequently dragged by the nearly horizontal magnetic field toward point B, producing a redshift in projection before returning to the solar interior. At the A-end, the GST/TiO intensities are generally high, indicating relatively hot plasma, while the Doppler velocities are close to zero compared to the strong redshifts observed toward the B-end.

Typically, the head regions of penumbral filaments exhibit significant Doppler blueshifts. Therefore, a strong blueshift is expected at the A-end. However, the inclination angles of the magnetic fields at A-end are nearly 90 degrees (horizontal), as seen in Figure~\ref{figure:fig4}(c4), resulting in the absence of blueshift. The nearly horizontal magnetic fields are partly due to projection effects, since the target region was located at a heliocentric angle of 30 degrees, and partly due to the stretching of the magnetic fields resulting from the shearing motions of umbrae. Although projection effects are known to become particularly severe at heliocentric angles larger than about 45 degrees for measurements of magnetic flux and other properties \citep{Falconer2016ApJ...833L..31F}, projection effects at smaller angles, such as the  $\sim$30 degrees considered here, can still significantly influence the observed Doppler signals. In particular, if the filament is oriented toward the northern direction, the head of the filament may also be displaced preferentially toward the north, thereby reducing the line-of-sight component of the flow and weakening the projected blueshift signal. A typical mature sunspot to the west of the target $\delta$ sunspot is shown in Figure~\ref{figb1}, of which the expected significant Doppler shifts were also missing near the head of its penumbral filaments. The projection effects also impact the LOS velocities at B-end, where significant redshifts are present. The stretching of the magnetic fields in region R1 also lead to narrower widths (68 and 92 km in panels b1 and b2) compared to the typical widths of penumbral filaments.
Along the filament, as the position moves toward the B-end, the redshifts gradually increase while the intensities decrease, reaching their extreme values at the B-end. This pattern likely results from a combination of the projected horizontal outflow and downward plasma motion. The maximum velocity in region R1 reaches 10 km\,s$^{-1}$, corresponding to a LOS magnetic field strength of 2155 G. Additionally, localized brightening near the redshifted edge suggests possible associated heating. 

An opposite flow pattern is observed in the blueshifted region 2, as shown in Figure~\ref{figure:fig4}. Similarly, we select a representative filament segment as shown in Figure~\ref{figure:fig4}(a), labeled C–D, which is characterized by a relatively bright head, a weaker main body, and a brighter bump at the tail in Figure~\ref{figure:fig4}(c1) and (c2). Plasma is observed to rise near point C and then flow horizontally along the filament toward point D. Along the filament, this horizontal flow is manifested as a blueshift toward the D-end, followed by a redshift signal associated with the bright tail, as shown in Figure~\ref{figure:fig4}(c1), (c2) and (c3). Same as A-end in region R1, the head (C-end) does not exhibit significant blueshifts. As shown in the Figure~\ref{figure:fig4}, the grains near C-end exhibit motions toward the sunspot umbra. Inward motions of penumbral grains have been reported previously and are commonly interpreted in terms of inclination differences between the filament and the surrounding magnetic field \citep{PGzhang2013A&A...560A..77Z,PGSobotka2022A&A...662A..13S}. Given that in the present case, the inward motion is directed northward (i.e., away from the solar disk center), and the target region is located at a heliocentric angle of around 30 degrees, the motions result in redshifts, which may counteract the expected blueshifts. Besides, the observation direction with a heliocentric angle of around 30 degrees can also reduce the blueshifts of the filamentary heads. The absence of the blueshifts can also be seen near the head regions of the mature sunspot shown in Figure~\ref{figb1}.

Therefore, if we classify the BLB structure as consisting of penumbral filaments, this interpretation is supported not only by its intermediate intensity between the umbra and the quiet-Sun regions and the presence of a nearly horizontal magnetic field, but also by its morphological similarity to standard penumbral filaments described by \citet{Tiwari2013ATiwari&A...557A..25T}. In the magnetoconvective picture, a penumbral filament can be regarded as a convective cell characterized by a bright head, a weaker main body, and a brighter tail, which is consistent with the filamentary properties observed in the BLB. The absence of significant Doppler velocities near the heads of the filaments is partly due to projection effects. A mature sunspot at the same heliocentric angle shown in Figure~\ref{figb1} also presents the absence of significant Doppler signals near the filamentary heads. In addition, the motions of the penumbral grains also counteract the expected Doppler signal. Taken together, these observational characteristics support an interpretation of the BLB structure in terms of penumbral filaments. It should be noted, however, that the BLB structure appears to be highly compressed compared to a classical, isolated penumbral filament. In such a crowded environment, interactions among neighboring filaments are likely to influence their individual motions and observational signatures. Therefore, not all filamentary structures within the BLB can be regarded as fully developed, standard penumbral filaments. 

Additionally, we find clear plasma flows along the filaments in the redshifted region 1 (Figure~\ref{figure:fig3}), where the flow is directed from A to B, that is, from the southern part toward the northern part of the structure. This provides strong evidence for field-aligned flows, consistent with the Evershed flow originating from the southern umbral core.

In the blueshifted region, prominent material flows are observed in the right segment of the structure, where the filaments are more nearly parallel to the PIL. In this case, the flow is directed southward, which is likewise consistent with an Evershed flow originating from the northern umbral core.
Here, we find that the orientation of the filaments significantly affects the observed horizontal plasma flows. In the central part of the structure, where the filaments are more nearly perpendicular to the PIL, the horizontal plasma flow is more difficult to detect. This may be due to the fact that the filaments in this region are relatively shorter compared to those in the adjacent regions.

In addition to the evidence discussed above, we further examine whether the magnitude of the projected velocities associated with the Evershed flow is consistent with expectations. As shown in Figure~\ref{figb1}, a simple sunspot located to the west of the $\delta$ sunspot that harbors the BLB, with a circular penumbra, exhibits the typical projection pattern of the Evershed flow, characterized by redshifts in the northern part and blueshifts in the southern part. The line-of-sight velocities ($V_{\mathrm{LOS}}$), in terms of their absolute values, are mostly about 1–2 km\,s$^{-1}$ in both the BLB and the adjacent western penumbra. If these are indeed projections of the Evershed flow, the corresponding horizontal velocities would be approximately 3–4 km\,s$^{-1}$ at this latitude, consistent with typical Evershed flow speeds.

Thus, taking into account the morphological resemblance between the filamentary structures in the BLB and penumbral filaments, the filament-aligned horizontal plasma flows consistent with Evershed flows from different umbral cores, and the consistency with typical Evershed flow speeds, the adjacent red- and blueshifted Doppler patches are most plausibly interpreted as signatures of Evershed flows along penumbral filaments associated with opposing umbral cores. Although such a scenario has been proposed previously \citep{Lites2002,Jiayi2023ApJ...955...40L,Cristaldi2014,Zuccarello2014}, clear observational evidence has so far been scarce. Our results provide high-resolution observational evidence of the flow pattern and strengthen this interpretation.

It is worth noting that the filaments exhibit multiple strands even within regions characterized by similar LOS velocities. As shown in Figure~\ref{figure:fig2}(e), within the upflow region, some filaments are oriented more vertically toward the umbra, while those on the western side tend to be more horizontal. The eastern part of the main downflow region exhibits a complex pattern, where localized upflows are embedded within areas of downflow. Moreover, the strongest magnetic fields are observed at the edges of both the upflow and downflow regions as shown in Figure~\ref{figure:fig2}(e). In the downflow region, the strongest magnetic field appears near the termination point of the Evershed flow. Similarly, in the upflow region, the strongest magnetic field is also located near the end of the Evershed flow. These features are more pronounced in the nearly parallel filaments, likely due to the enhanced shear structures present there.

Based on the high-resolution GST observations and long-duration SDO/HMI monitoring, we propose a scenario that explains the formation of the BLB. As illustrated in Figure~\ref{figure:fig7}, opposite polarities from two sunspot pairs (P1–N2) converge during emergence, with each developing its own penumbra. 
When the two sunspots come into close contact, their penumbrae interlace, forming the observed BLB. We examined the stage before the convergence of P1 and N2 (Figure~\ref{figure:fig6}). At that time, the sunspot P1 had already developed its penumbra. 
The blueshifted flow in its penumbra is consistent with the projected Evershed flow, since the target region was located in the northwest of the solar disk. 
The persistence of the flows in the penumbra, the intermediate intensity between those of umbrae and quiet Sun regions, and the strongly inclined magnetic field during the formation of the BLB suggest that the BLB results from interlace of penumbrae.

Moreover, the fine structure of the BLB is very similar to a photospheric structure called "orphan penumbra" \citep{Jur2014A&A...564A..91J, Zuccarello2014}.
Both structures form above PILs, and are dominated by horizontal magnetic fields.
Large-scale adjacent red and blueshifted patches Doppler shift patterns are observed in these structures. 
Besides, similar to the flux ropes lying above orphan penumbrae \citep{Zuccarello2014}, the flux overlying the BLB, manifested as the fibrils seen in $\mathrm{H\alpha}$ image, is also evident in our observations. 
These similarities also suggest that the BLB is intrinsically formed by the evolution of the penumbrae.
The noteworthy distinction between the BLB and orphan penumbrae lies in the surrounding vertical magnetic field regions. 
For the orphan penumbrae, the vertical magnetic field drops below $\sim$1.8 kG \citep{Jur2018A&A...611L...4J}, and the opposite-polarity pores evolve into penumbrae without umbral cores. 
By contrast, the field strength is much stronger during the formation of the BLB, and the umbrae are preserved.
The strong surrounding fields associated with BLBs enable greater magnetic energy storage, thereby endowing them with a higher eruption potential.

The BLB exhibited remarkable stability during the entire GST observational period. 
The $\mathrm{H\alpha}$ line center image in Figure~\ref{figure:fig2}(b) reveals large-scale chromospheric arches overlying the BLB. The post-flare loops were also generated through magnetic reconnection during the flare, as seen in the SDO/AIA EUV observations shown in Figure B.2. These arching magnetic structures seen in the Halpha and EUV bands may play a role in confining the BLB, contributing to its stability during the 5.5-hour GST observation.

During the formation of the BLB, the shearing and rotational motions of the sunspots progressively stretched the intervening horizontal magnetic fields, and further contributed to the superstrong horizontal magnetic field in the BLB \citep{Hotta2020MN}. 
This process led to the gradual elongation and fragmentation of the BLB, ultimately triggering complex eruptive events, which is consistent with the collisional shearing scenario proposed by \citet{Yan2012AJ....143...56Y}, \citet{Chintzoglou2019}, and \citet{Boocock2020ApJ...900...65B}. 
In this paradigm, PILs form where opposite polarities from distinct bipoles converge, driving shearing motions, flux cancellation, and sequences of solar flares with associated coronal mass ejections (CMEs), which is a fundamental driver of intense solar activity \citep{Chintzoglou2019, Liu2019ApJ...884...45L}. 
Under the influence of strong shearing and converging of horizontal magnetic flux—induced by sunspot proper motions—magnetic helicity is continuously injected, fostering the formation of increasingly complex magnetic flux ropes that eventually erupt. 
The physical nature of collisional shearing regions remains incompletely understood due to the observational limitations. 
Our high-resolution observations provide a more detailed physical insight that advances the current understanding.

\section{Summary and conclusions}
\label{sect:conclusion}

In this work, we present a detailed analysis of the observation of a BLB in NOAA AR 13663. 
Unlike the traditional light bridges, high-resolution observations in the TiO band obtained by GST reveal that the BLB is composed of filamentary segments, each exhibiting different orientations. 
The BLB hosts strongly inclined magnetic field with a strength exceeding 4000~G, consistent with previous studies \citep{Wang2018,Castellanos2020ApJ...895..129C,Castellanos2025}.
The longitudinal velocity map shows adjacent red and blueshifted patches flow along the BLB. 
The filaments in the downflow region (R1) are longer and narrower than those in the upper flow region (R2). 
The BLB remains stable throughout the 5.5-hour GST observation period. 
Fibril structures, overlapping the BLB, are observed in the H$\alpha$ band, likely being the reason for the stability. 

The formation of the $\delta$-type sunspot associated with the BLB was captured by HMI, which records the emergence of a new sunspot pair (N2–P2) near a pre-existing pair (N1–P1). 
The converging and shearing motions of the sunspots N2 and P1 lead to the $\delta$-type sunspot and the formation of the BLB.
HMI Doppler maps of the BLB region show similar adjacent red and blueshifted patches patterns, though at lower resolution, with the signals pre-existing in the penumbrae of the sunspots N2 and P1 before the BLB formation. 
These pre-existing velocity patterns are likely caused by the Evershed flows in the penumbrae of the two sunspots.

Based on these observational evidences, we proposed that the BLB is the compressed and stretched penumbrae of two interacting sunspots. 
The filamentary segments with opposite Doppler shifts belong to distinct penumbrae, consistent with previous MHD simulations explaining the strong magnetic fields in BLBs \citep{Hotta2020MN}. 
The BLB is confined by the fibrils seen in the $\mathrm{H\alpha}$ band and the Post-flare loops seen in the EUV bands shown in Figure \ref{figb2}. The strongest magnetic fields appear near the ends of the Evershed flows and in the more sheared penumbral filaments. 

The BLB shares similarities with the so-called orphan penumbrae, which are dominated by inclined fields and stabilized by overlying flux \citep{Jur2014A&A...564A..91J, CastellanosDuran2025A&A}. 
The similarity between these two structures highlights the intricate interconnectedness of solar phenomena. 
Nevertheless, several fundamental questions remain open for investigation. 
Such as how such BLBs evolve to large-scale magnetic flux ropes in the corona and contribute to eruptive events \citep{Chen2022ApJ...937...91C}. 
Future coronal magnetic field measurements will be critical for addressing this coupling between BLBs and the upper solar atmosphere \citep{yang2020Sci...369..694Y, Yang2024}.

\begin{acknowledgments}
This work is supported by the National Natural Science Foundation of China (grants 12425301, 12473051, and 12403065) and China's Space Origins Exploration Program.
We appreciate the data support from both the GST and SDO/HMI teams. 
The Goode Solar Telescope (GST) at the Big Bear Solar Observatory (BBSO) is operated by the New Jersey Institute of Technology with support from the U.S. National Science Foundation (NSF; grant AGS-2309939), and is partially funded by the Korea Astronomy and Space Science Institute and Seoul National University. 
We also thank Professor Hui Tian from Peking University for helpful discussions.
\end{acknowledgments} 

\appendix

\counterwithin{figure}{section}

\section{Milne-Eddington inversion}\label{ME}

A Milne-Eddington inversion code was applied to the spectropolarimetric observations of the \ion{Fe}{1} line obtained by GST/NRIS. 
This inversion code\footnote{Available at \url{https://doi.org/10.5281/zenodo.18397378}}, written in C and paralleled with MPI, follows the analytic solutions described in \citet{delToroIniesta2003} and \citet{LL04}. 
In this framework, the direction of positive Stokes $U$ corresponds to a counterclockwise rotation of 45$^\circ$ with respect to that of positive Stokes $Q$, while Stokes $V$ represents the difference between the right- and left-handed circular polarization. 
The observer is supposed to be facing the radiation source.
The inverted inclination angle ($\theta_B$) is defined as the angle between the magnetic field vector and the direction pointing to the observer, and the azimuth angle ($\phi_B$) is the counterclockwise from direction of positive Stokes $Q$. 
Following \citet{Li2019ApJ}, the code adopts the sum ($S=S_0+S_1$) and a ratio $\beta = S_0/(S_0+S_1)$ instead of the source function ($S_0$) and its gradient ($S_1$), since $S$ corresponds to the continuum intensity, which is uncorrelated with other parameters. 
In addition, a folded boundary condition was applied to $\phi_B$ to prevent solutions from becoming trapped at the bounds. An inversion test with Hinode/SOT-SP \citep{Tsuneta2008SoPh, Ichimoto2008SoPh} observations is shown in Figures~\ref{figb2}.  
Only 50 wavelength grids across the \ion{Fe}{1} 6302.5~{\AA} are used in this code. 
Top panels show $B$ (magnetic field strength), $\theta_B$ and $\phi_B$ inferred with the MERLIN code \citep{Lites2007MmSAI}, while the bottom panels show the results from this code. 
Since the filling factor is not considered in this code, $B$ from the MERLIN code shown is multiplied by the filling factor. 
Overall, the two codes yield consistent inversion results. 
\begin{figure*}[htp]
  \center
  \includegraphics[width=\textwidth]{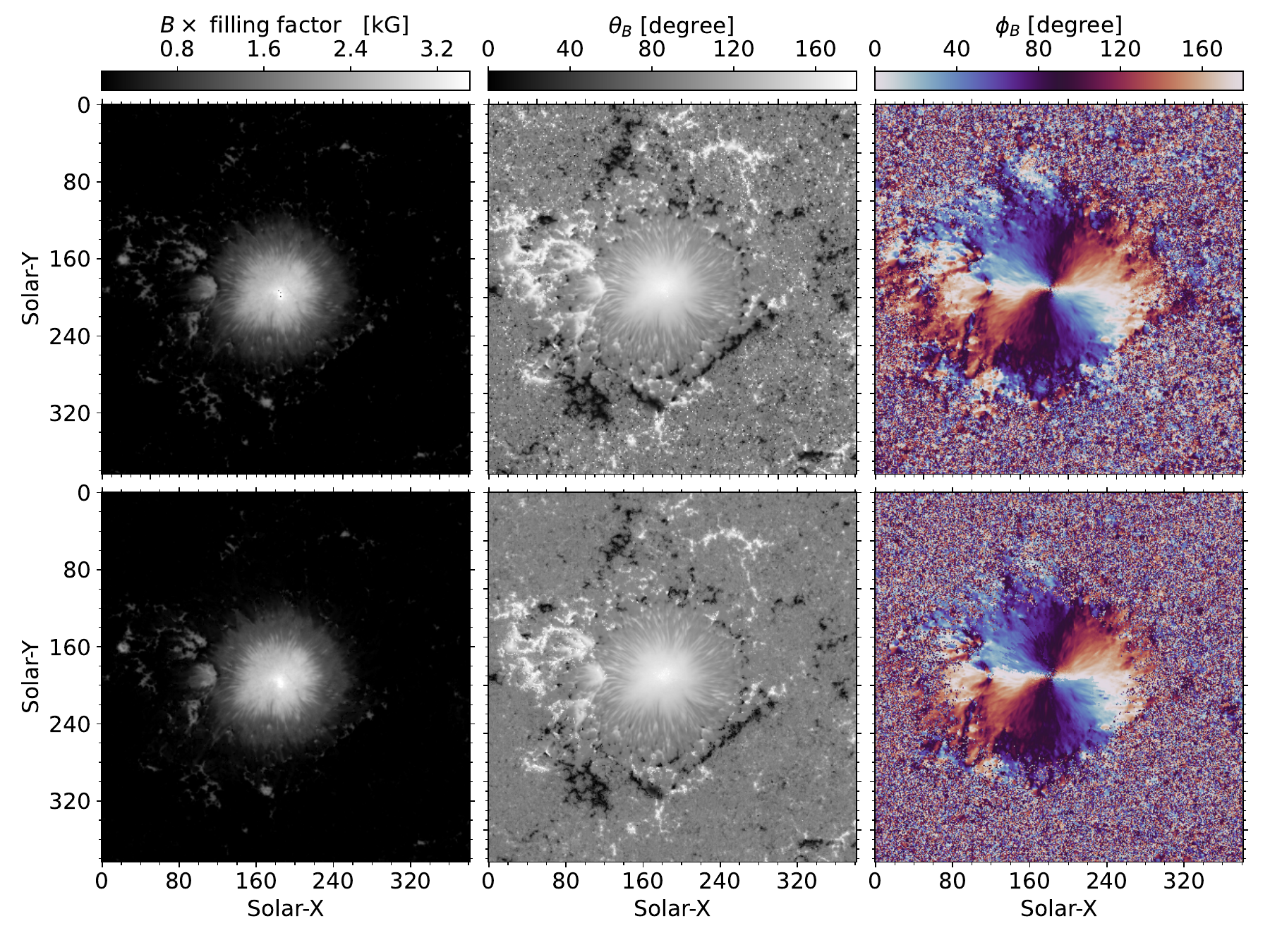}
  \caption{Inversion test with the Hinode/SOT-SP observations. Top and bottom panels show the results from MERLIN and this code, respectively.
  \label{figa1}}
\end{figure*}

\section{Supplementary figures}\label{Supfigures}
This section presents supplementary figures that complement the discussions in Section~\ref{sect:discussion}. The details of the figures shown in this section are discussed in the main text. 

\begin{figure*}[htp]
  \center
  \includegraphics[width=\textwidth]{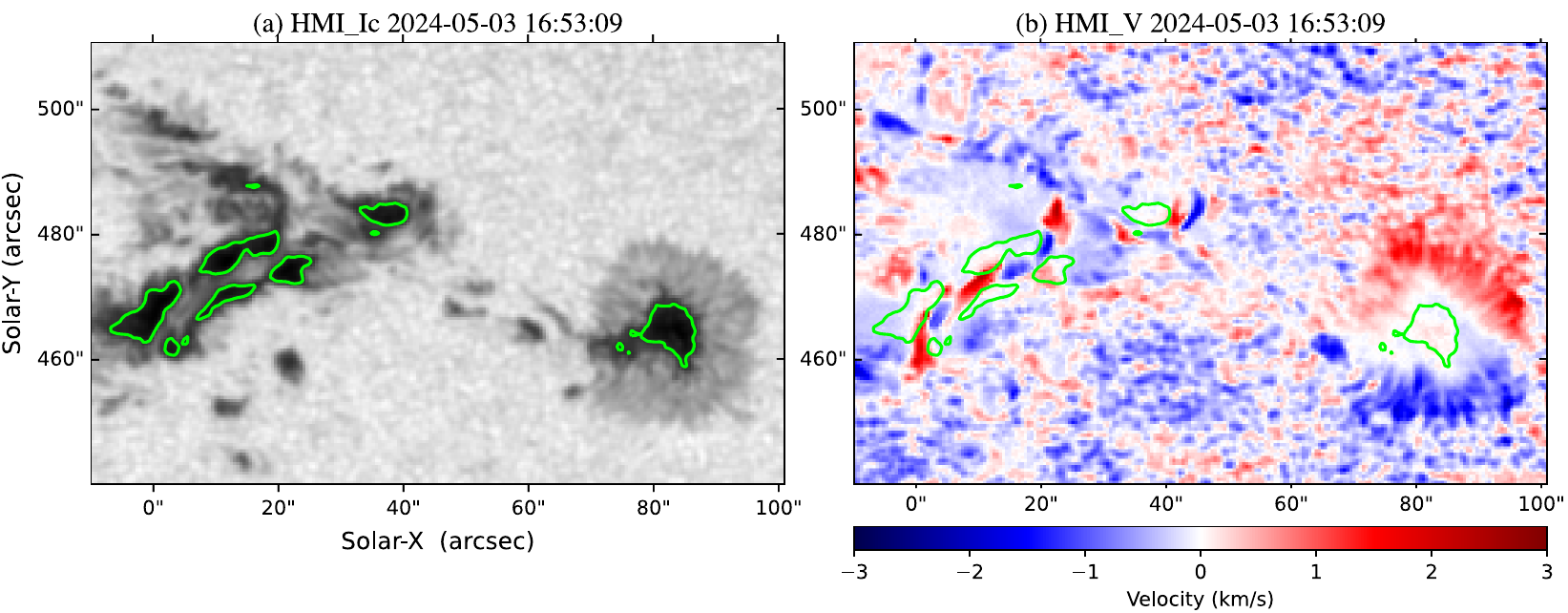}
  \caption{Large field-of-view image showing a simple sunspot near the BLB that displays the typical projection of the Evershed flow. 
  The lime contours denote the 0.35 \,$I_{c}$.
  \label{figb1}}
\end{figure*}
\begin{figure*}[htp]
  \center
\begin{interactive}{animation}{animation7.mp4}
\includegraphics[width=1\textwidth]{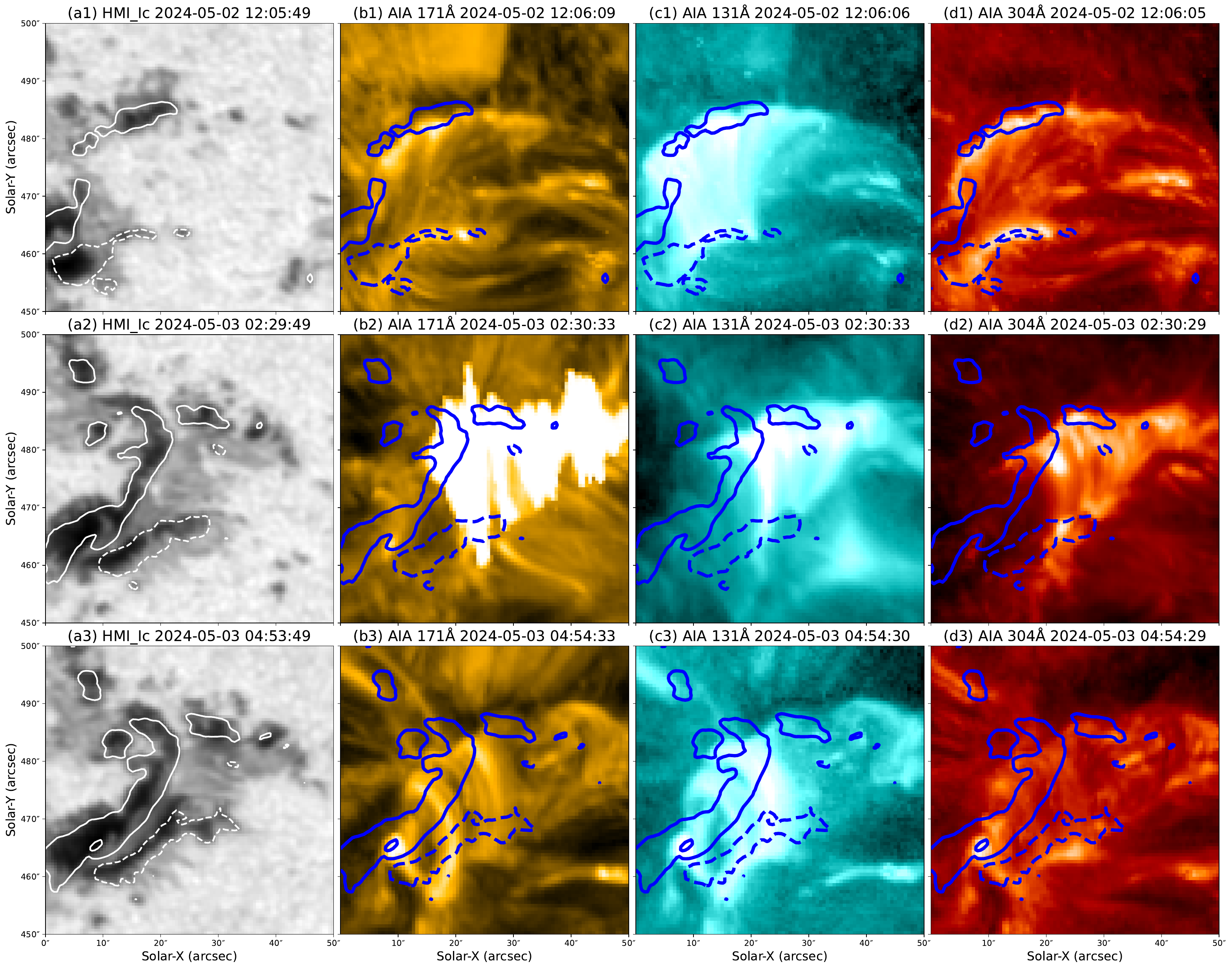}
\end{interactive}
  \caption{Post-flare loops associated with flare activities in the upper solar atmosphere are observed in the EUV channels above the final BLB. The FOV is referenced to the coordinates at the time of Figure~\ref{figure:fig1}. 
  Solid white and blue contours represent +1000 Gauss, while dashed white and blue contours represent -900 Gauss.
  An animation is available showing the evolution of EUV images associated with flare activity. The animation runs from May 2nd, 2024 11:00:06 to May 5th, 2024 13:48:30. The animation's real-time duration of 4 seconds.
  \label{figb2}}
\end{figure*}

\bibliography{bibli}{}
\bibliographystyle{aasjournalv7}

\end{document}